\documentclass[5p,times,twocolumn]{elsarticle}

\usepackage{amssymb}
\usepackage{booktabs}
\usepackage{adjustbox}
\usepackage{siunitx}
\usepackage{lscape}
\usepackage{multirow}
\usepackage{tikz}
\usepackage{colortbl}
\usepackage{array}
\usepackage{svg}
\usepackage{graphicx} 
\usepackage[unicode,breaklinks=true]{hyperref}
\hypersetup{
	colorlinks=true,
   	linkcolor=blue,
   	citecolor=blue,
    filecolor=blue,
   	urlcolor=blue}
\usepackage{colortbl}
\usepackage{hhline}
\usepackage{pdflscape}
\usepackage{caption} 
\usepackage{graphicx,calc}
\usepackage{makecell}

\newcolumntype{L}[1]{>{\raggedright\let\newline\\\arraybackslash\hspace{0pt}}m{#1}}
\newcolumntype{C}[1]{>{\centering\let\newline\\\arraybackslash\hspace{0pt}}m{#1}}
\newcolumntype{R}[1]{>{\raggedleft\let\newline\\\arraybackslash\hspace{0pt}}m{#1}}

\usepackage{soul}
\usepackage{color}
    \definecolor{celadon}{rgb}{0.67, 0.88, 0.69}
    \definecolor{flamingopink}{rgb}{0.99, 0.56, 0.67}
    \definecolor{lovelygreen}{rgb}{0.54, 0.90, 0.60}

\usepackage[figuresright]{rotating}
\usepackage{pbox}
\usepackage{makecell}
\usepackage{appendix}

\usepackage{footnote}
\usepackage{tablefootnote}

\usepackage{graphicx}
\usepackage{tabularray}
\usepackage{xcolor}

\definecolor{gray7}{HTML}{BFBFBF}
\definecolor{gray9}{HTML}{E5E5E5}

\usepackage{tabularray}

\makesavenoteenv{tabular}
\makesavenoteenv{table}

\journal{DFRWS EU 2026}

\begin{document}
\emergencystretch 3em

\begin{frontmatter}




\title{AutoDFBench 1.0: A Benchmarking Framework for\\Digital Forensic Tool Testing and Generated Code Evaluation}

\renewcommand{\theaffn}{\arabic{affn}}


\author[label1]{Akila Wickramasekara}
\author[label2]{Tharusha Mihiranga}
\author[label3]{Aruna Withanage}
\author[label4]{Buddhima Weerasinghe}
\author[label3]{Frank Breitinger}
\author[label2]{John Sheppard}
\author[label1]{Mark Scanlon}

\affiliation[label1]{organization={Forensics and Security Research Group, School of Computer Science},
            addressline={University College Dublin},
            city={Belfield},
            state={Dublin 4},
            country={Ireland}}

\affiliation[label2]{organization={Forensics and Security Research Group, Department of Computing \& Mathematics},
            addressline={South East Technological University},
            country={Ireland}}

\affiliation[label3]{organization={Chair for Cybersecurity},
            addressline={University of Augsburg},
            city={Augsburg},
            country={Germany}}

\affiliation[label4]{organization={School of Computer Science},
            addressline={University of Birmingham},
            city={Birmingham},
            country={United Kingdom}}




    

\begin{abstract}

The National Institute of Standards and Technology (NIST) Computer Forensic Tool Testing (CFTT) programme has become the \textit{de facto} standard for providing digital forensic tool testing and validation. However to date, no comprehensive framework exists to automate benchmarking across the diverse forensic tasks included in the programme. This gap results in inconsistent validation, challenges in comparing tools, and limited validation reproducibility. This paper introduces AutoDFBench 1.0, a modular benchmarking framework that supports the evaluation of both conventional DF tools and scripts, as well as AI-generated code and agentic approaches. The framework integrates five areas defined by the CFTT programme: string search, deleted file recovery, file carving, Windows registry recovery, and SQLite data recovery. AutoDFBench 1.0 includes ground truth data comprising of 63 test cases and 10,968 unique test scenarios, and execute evaluations through a RESTful API that produces structured JSON outputs with standardised metrics, including precision, recall, and F1~score for each test case, and the average of these F1~scores becomes the \emph{AutoDFBench Score}. The benchmarking framework is validated against CFTT's datasets. The framework enables fair and reproducible comparison across tools and forensic scripts, establishing the first unified, automated, and extensible benchmarking framework for digital forensic tool testing and validation. AutoDFBench 1.0 supports tool vendors, researchers, practitioners, and standardisation bodies by facilitating transparent, reproducible, and comparable assessments of DF technologies.  
\end{abstract}




\end{frontmatter}












\section{Introduction}
\label{intro}

The rapid growth of digital crimes, accelerated by the emergence of generative Artificial Intelligence (AI) technologies, has significantly increased the workload and complexity faced by digital forensic (DF) investigators~\cite{WICKRAMASEKARA2025301859, Costantini2019}. To address these challenges, there is a growing demand for robust, accurate, and consistent digital forensic tools that can ensure reliability and reproducibility in evidence analysis. However, one of the most persistent issues in this domain remains the lack of standardisation in tool validation and evaluation methodologies~\citep{9720948}.

Recent advancements in large language models (LLMs) and autonomous AI agents have shown great potential in enhancing the efficiency and productivity of digital investigations~\citep{Wickramasekara2024Framework, 2023SCANLONChatGPT,11126744}. These systems can generate digital forensic scripts, automate repetitive tasks, and assist in interpreting complex data. Nevertheless, despite their promise and growing interest from the digital forensic research community~\cite{Breitinger202410Year}, there is still no appropriate or standardised mechanism to validate and benchmark the accuracy, reliability, and forensic soundness of AI-generated source code and scripts.

This research addresses that gap by focusing on the significant enhancement and expansion of a comprehensive unified benchmarking framework for evaluating both conventional and AI-generated digital forensic tools and scripts, AutoDFBench. AutoDFBench builds upon the principles of the National Institute of Standards and Technology (NIST) Computer Forensics Tool Testing (CFTT) programme and provides an automated, extensible, and standardised mechanism for reproducibly evaluating forensic tools across multiple investigation scenarios. This paper advances the framework from an initial proof-of-concept to a 1.0 release version, covering the breadth of CFTT software test suites needed for computer forensic software tool testing and validation.

This work makes the following contributions:

\begin{enumerate}
    \item Enhance and expand the AutoDFBench framework through the incorporation of an improved API layer, a new CSV input layer, and the expansion of the set of test suites from one to five. The framework is released open source under the Apache 2.0 License via the project's GitHub repository (\url{https://github.com/akila-UCD/AutoDFBench}).

    \item The preparation and integration of comprehensive ground truth datasets derived from the NIST CFTT programme, encompassing 63 test cases and 10,968 unique test scenarios across five test suites.

    \item The experimental evaluation of the framework using NIST provided datasets demonstrates its reliability, accuracy, and consistency in benchmarking digital forensic tools and AI-generated code.
\end{enumerate}

\begin{table}[!t]
\centering
\small
\caption{Forensic String Search Test Cases}
\label{tab:desc-string-search}
\resizebox{\columnwidth}{!}{%

\begin{tabular}{|l|p{6.8cm}|}
\hline
\textbf{Test Case} & \textbf{Description} \\
\hline
FT-SS-01 & Test the ability to identify a plain ASCII string consisting of a single word. \\
\hline
FT-SS-02 & Assess whether a tool can locate an ASCII string regardless of upper or lowercase variations. \\
\hline
FT-SS-03 & Verify that a tool matches whole words exactly and does not detect substrings, and ignores case. \\
\hline
FT-SS-04 & Evaluate searches that require two different terms to be present within the same file. \\
\hline
FT-SS-05 & Check if a tool can detect the presence of at least one string from a given set of words. \\
\hline
FT-SS-06 & Confirm whether a search can identify a string while excluding content containing another specified term. \\
\hline
FT-SS-07 & Determine support for searching text in non-English scripts, including Asian characters, Hangul, Kana, Cyrillic, Latin with accents, and right-to-left languages. \\
\hline
FT-SS-08 & Assess built-in search functions for structured data types such as email addresses, phone numbers, or identification numbers. \\
\hline
FT-SS-09 & Measure performance in scenarios such as formatted documents, fragmented files, inaccessible storage areas, NTFS metadata, substrings in file names, and word stemming. \\
\hline
FT-SS-10 & Examine the handling of pattern-based searches using regular expressions, including hexadecimal matching and simple character sets. \\
\hline
\end{tabular}
}

\end{table}


\section{Computer Forensics Tool Testing Programme}
\label{bg_cftt}

The CFTT programme\footnote{\url{https://www.nist.gov/itl/ssd/software-quality-group/computer-forensics-tool-testing-program-cftt}} provides a structured methodology for evaluating digital forensic tools, aiming to ensure reliable, repeatable, and standardised results that withstand legal scrutiny. It includes formal test specifications and publishes tool testing results that assess performance across tasks, including string search, file carving, deleted file recovery, registry analysis, database extraction, write blocking, cloud data extraction, media preparation, and mobile devices. These suites are widely recognised as \textit{de facto} standards for independent DF tool validation. In this work, CFTT underpins AutoDFBench~1.0, supplying the benchmark datasets and specifications required for automated, reproducible, and comparable evaluations of forensic tools and scripts.

\subsection{Forensic String Search (FSS)}
\label{intro-fss}

In FSS, NIST emphasises two main requirements for a string search tool. One requirement is that the tool returns exact matches based on the given query, and the other is that it should support at least one character representation for searching.
To address these requirements, forensic string search is accompanied by ten primary test cases defined by the NIST CFTT programme, which are explained in Table~\ref{tab:desc-string-search}. 

These ten test cases are subdivided into sub-test cases, as each test case consists of different string values, file statuses (active, deleted, and unallocated), and the disk operating system.

\begin{table}[!t]
\centering
\small
\caption{Deleted File Recovery Core Test Cases}
\label{tab:dfr_testcases}
\resizebox{\columnwidth}{!}{%

\begin{tabular}{|c|p{6.8cm}|}
\hline
\textbf{Test Case} & \textbf{Description} \\ \hline
DFR-01 & Evaluates the recovery of a single non-fragmented file. \\ \hline
DFR-02 & Assesses the recovery of a file consisting of two fragments. \\ \hline
DFR-03 & Examines the recovery of a file that has multiple fragments. \\ \hline
DFR-04 & Tests recovery of several non-fragmented files with non-Latin character filenames. \\ \hline
DFR-05 & Evaluates the recovery of two fragmented files. \\ \hline
DFR-06 & Check the recoverability of a single large file. \\ \hline
DFR-07 & Tests recovery of one file that has been overwritten. \\ \hline
DFR-08 & Examines the recovery of multiple overwritten files. \\ \hline
DFR-09 & Evaluates recovery when a large number of files are deleted without overwriting. \\ \hline
DFR-10 & Assesses recovery of a large number of files that were deleted with partial overwriting. \\ \hline
DFR-11 & Tests recovery of files deleted from a single directory. \\ \hline
DFR-12 & Assesses recovery of files deleted across multiple directories. \\ \hline
DFR-13 & Checks recovery of chaotic file system activity. \\ \hline
DFR-14 & Evaluates recovery of non-standard or other file system objects. \\ \hline

\end{tabular}
}
\end{table}

\subsection{Deleted File Recovery}

Recovery of file system metadata is a crucial task in digital forensics, and tools that support this exercise are subject to mandatory requirements defined by the NIST CFTT programme. These tools should correctly identify deleted entries and deleted file system metadata. They should report errors and be able to construct recovered objects from allocated or non-allocated space, consisting of data blocks from deleted blocks.

Considering these requirements, NIST defined 14 mandatory test cases and three optional test cases (DFR-15, DFR-16, DFR-17). Table~\ref{tab:dfr_testcases} summarises the mandatory test cases.

\begin{table}[!t]
\centering
\small
\caption{File Carving Test Cases}
\label{tab:file_carving_cases}
\resizebox{\columnwidth}{!}{%

\begin{tabular}{|c|p{6.8cm}|}
\hline
\textbf{Test Case} & \textbf{Description} \\ \hline
FC-01 & Evaluates recovery of sector-aligned contiguous files with no padding between files. \\ \hline
FC-02 & Assesses recovery of byte alligned contiguous files without padding between files. \\ \hline
FC-03 & Tests recovery of contiguous files with padding added between files. \\ \hline
FC-04 & Evaluates recovery of files with fragmented order. \\ \hline
FC-05 & Check the recovery of files with out of order fragments. \\ \hline
FC-06 & Assesses recovery of an intermixed fragmented file. \\ \hline
FC-07 & Tests recovery of partial or incomplete files. \\ \hline
\end{tabular}
}
\end{table}

\begin{table}[!t]
\centering
\small
\caption{Windows Registry Core Feature Test Cases}
\label{tab:registry_testcases}
\resizebox{\columnwidth}{!}{%

\begin{tabular}{|c|p{6.5cm}|}
\hline
\textbf{Test Case} & \textbf{Description} \\ \hline
NR-01 & Verify support for processing different data types. \\ \hline
NR-02 & Evaluate registry file with simple tree structure \\ \hline
NR-03 & Evaluate registry trees with 512 levels or more levels. \\ \hline
NR-04 & Assess files which have long key names.(255 or more bytes) \\ \hline
NR-05 & Assess files which have long value names. (16,383 or more bytes) \\ \hline
NR-06 & Evaluate handling file with large data. (> 16,344 bytes) \\ \hline
NR-07 & Verify correct handling of non-ASCII characters. \\ \hline
NR-08 & Test the scenarios of unusual names for keys and values. \\ \hline
CR-01 to CR-05 & Validate tool behaviour when processing corrupted or wiped hive files. \\ \hline
MR-01 to MR-15 & Assess resilience against manipulated hive files (e.g., hidden keys, hidden values, invalid data sizes, ambiguous encodings). \\ \hline
\end{tabular}
}
\end{table}

\subsection{File Carving}

File carving is an essential technique in digital forensics, reconstructing files from internal structures without original metadata, and it is most frequently used to recover artefacts from unallocated space~\cite{4806206}. To evaluate the file carving capabilities in DF tools, NIST created a set of test cases covering a variety of scenarios, including contiguous, padding-added, and byte-shifted files. These test cases are detailed in Table~\ref {tab:file_carving_cases}. Along with these test cases, they also provided test disk image files containing example data related to each test scenario.

\subsection{Windows Registry Recovery}

The Windows Registry is a central database that records system and configuration settings, user activity, and application settings, making it a critical source of forensic evidence~\cite{Singh03052020}. NIST provided three main requirements for a Windows registry recovery tool. One tool should support one or more hive files and a disk image with a Windows partition. The second requirement is that the tool should be able to notify users about abnormal information in the hive registry files. Lastly, it should be able to interpret the registry objects. With these requirements, NIST defined 15 core test cases and 25 optional test cases to evaluate a registry recovery tool. Table~\ref{tab:registry_testcases} summarises the core test cases for Windows Registry extraction. Normal Registry (NR) test cases focus on the structure of the hive file, while Corrupted Registry (CR) test cases evaluate the tool's ability to recover corrupted hive files. Manipulated Registry (MR) test cases assess the tool's reaction to manipulated hive files.

\subsection{SQLite Data Recovery}

SQLite databases are widely used across operating systems, applications, and mobile devices, making them an essential source of digital forensic evidence. Data recovery from SQLite is particularly challenging because deleted or updated rows may persist in unallocated pages, freelists, or database journaling files such as the \texttt{journal} or Write-Ahead Log (WAL)~\cite{MENG2019S31}.

To provide a standard for evaluating tool reliability, the NIST CFTT programme defines five core features that are required of a SQLite recovery tool. These requirements include no file modification for the analysed file, database configuration recoverability, database schema recoverability, table content recoverability, and the recoverability of the source of data elements. 

To cater to these requirements, NIST defined four test cases, which are summarised in Table~\ref{tab:sqlite_testcases}.

\begin{table}[h!]
\centering
\small
\caption{SQLite Data Recovery Core Test Cases}
\label{tab:sqlite_testcases}
\resizebox{\columnwidth}{!}{%

\begin{tabular}{|c|p{6.5cm}|}
\hline
\textbf{Test Case} & \textbf{Description} \\ \hline
SFT-01 & Verifies that the tool correctly recovers page size, journal mode, number of pages and text encoding.\\ \hline
SFT-02 & Ensures that the tool reports the complete schema, including table listings, column names, and row information for every table. \\ \hline
SFT-03 & Checks that the tool recovers and reports all rows, including deleted or updated entries. \\ \hline
SFT-04 & Verifies that the tool reports the source file name for all recovered data elements. \\ \hline
\end{tabular}
}
\end{table}

\subsection{Motivation for this Work}
\label{motivation}

Existing validation efforts, such as those provided through the NIST CFTT programme, have laid important groundwork for standardising forensic tool testing. However, these results are typically published as descriptive classifications, which makes it challenging to compare tools directly or reproduce evaluations consistently. Furthermore, most evaluations require significant manual effort and lack an automated mechanism to generate precise performance metrics.  

This work is motivated by the need for a unified and automated benchmarking framework that can provide clear, quantifiable measures, such as precision, recall, and F1~score, across diverse forensic scenarios. By offering reproducible, metrics-driven evaluation, AutoDFBench 1.0 enables a more transparent comparison of tools and supports both practitioners and researchers in assessing the reliability of forensic outputs.

Keyword searching, deleted data recovery, and timeline analysis are three of the most commonly used techniques by digital forensic investigators \cite{hargreaves2024dfpulse}. Hence, AutoDFBench 1.0 focuses on implementing the corresponding and relevant NIST CFTT test suites on these topics. The String Search, Deleted File Recovery, and File Carving test suites correspond to the first two techniques -- keyword searching and deleted data recovery. The corresponding test suites relevant for timeline analysis are the Windows Registry Recovery and the SQLite Data Recovery.

\section{Related Work}
\label{RelatedWork}

A benchmark is defined as a tool for comparative evaluation according to a specific characteristic, and a proper benchmarking mechanism should inherit key attributes such as repeatability, reproducibility, relevance, fairness, verifiability, usability, and a properly controlled dataset~\cite{wfs21474, buildbenchmark}. \citet{wfs21474} discusses the distinction between validating a forensic method and validating a tool. At the same time, it is mentioned that the benchmark should include factors such as tool version, testing party, and the frequency of tests. Additionally, the testing lab must adhere to accreditation for tool testing, which indicates that the tool testing is conducted in a rigorous yet subjective manner.

In another study, \citet{10.1145/2016039.2016088} introduces a benchmark framework for mobile data acquisition. The study examined parameters such as the type of data acquired and the time required for data acquisition, which varied by device model and forensic tools. Yet, the authors conclude that they plan to include images, messages, and deleted data in the benchmark. \citet{PAN200971} introduced a performance testing benchmark for DF tools by utilising the execution time in encryption key recovery scenarios. Moreover, the authors mentioned that the NIST CFTT and the Scientific Working Group on Digital Evidence (SWGDE) focused on evaluating the correctness of digital forensics tools. Still, some of the data are not publicly available.

The ubiquitous DF acquisition, analysis, and reporting steps necessitate investigators to rely on bespoke DF hardware and software tools. The reliability and accuracy of these tools would be 100\% in an ideal scenario, but the field does not demand 100\% accuracy 100\% of the time~\cite{HORSMAN2019163}. Rather, it requires accurate reporting on their reliability and constraints. \citet{HORSMAN2019163} highlights benchmarking challenges, including an insufficient dataset for evaluation, which focuses solely on the tool's correct functionality rather than its proper usage and error identification. Additionally, the author states that NIST provides the closest thing to a comprehensive tool for testing; however, it still does not cover all areas. Consequently, the field currently has a gap in insufficient testing standards and procedures to effectively validate a tool. \citet{9678340} studied the existing challenges in digital forensics, focusing on the obstacles within the resources category. They highlighted a significant gap in the unavailability of benchmarks and datasets, which hinders proper evaluations.

\citet{cherif2025dfirmetricbenchmarkdatasetevaluating} proposes the DFIR-Metric to evaluate LLMs, which supports digital forensics investigation tasks. It comprises three modules: Multiple-Choice Questions (MCQs), capture the flag tasks, and disk analysis tasks, utilising NIST CFTT's forensic string search test suite. The authors evaluate the LLMs by categorising output as correct, skipped, syntax, and wrong, then calculating how many occurrences were accurate from the tested LLMs.

AutoDFBench has recently been introduced, which can evaluate tools and generate code from GenAI models focusing on the NIST forensic string search test suite~\cite{AkilaAutoDFBench}. Existing work on AutoDFBench only considers this single testing scenario, and the expansion of this framework forms part of the motivation for this work, as outlined in greater detail in Section~\ref{motivation}.

Despite the progress of these studies, a significant research gap remains. Existing validation frameworks often focus on narrow tool categories or rely on manual, descriptive assessments, limiting reproducibility and comparability. None provides a unified, automated, and comprehensive benchmarking of both conventional and AI-generated forensic tools and scripts. This gap directly impacts trust, transparency, and standardisation in digital forensic investigations -- precisely the challenge that \textit{AutoDFBench~1.0} seeks to address through complete, automated, score based, and reproducible benchmarking.

This work extends the AutoDFBench framework by adding additional NIST test suites, including deleted file recovery, file carving, Windows registry recovery, and SQLite recovery. It also adds an API layer, integration, and a CSV input layer -- elevating the framework's versioning to \textit{AutoDFBench~1.0}.

\subsection{AI-Generated Code in Digital Forensics}
\label{AIGenerated}

Recent advances in LLMs and domain-specific autonomous agents have introduced a new paradigm in digital forensic research.
These systems have the potential to automatically generate code fragments or scripts that replicate forensic functions for scenarios such as string search, file carving, deleted file recovery, etc.~\cite{2023SCANLONChatGPT, AkilaAutoDFBench}. Such generative capabilities offer opportunities for rapid prototyping, automated evidence analysis, and intelligent assistance during investigations. However, they also introduce new challenges related to the reliability, reproducibility, and forensic soundness of automatically generated code and scripts. As highlighted by \citet{DUNSIN2024301675}, although AI and machine learning techniques are increasingly integrated into digital forensics and incident response, they remain hindered by issues of data validity, interpretability, and standardised validation procedures. 

\section{Framework Design}
\label{framworkDesign}

\begin{figure}
    \centering
    \includegraphics[width=\columnwidth]{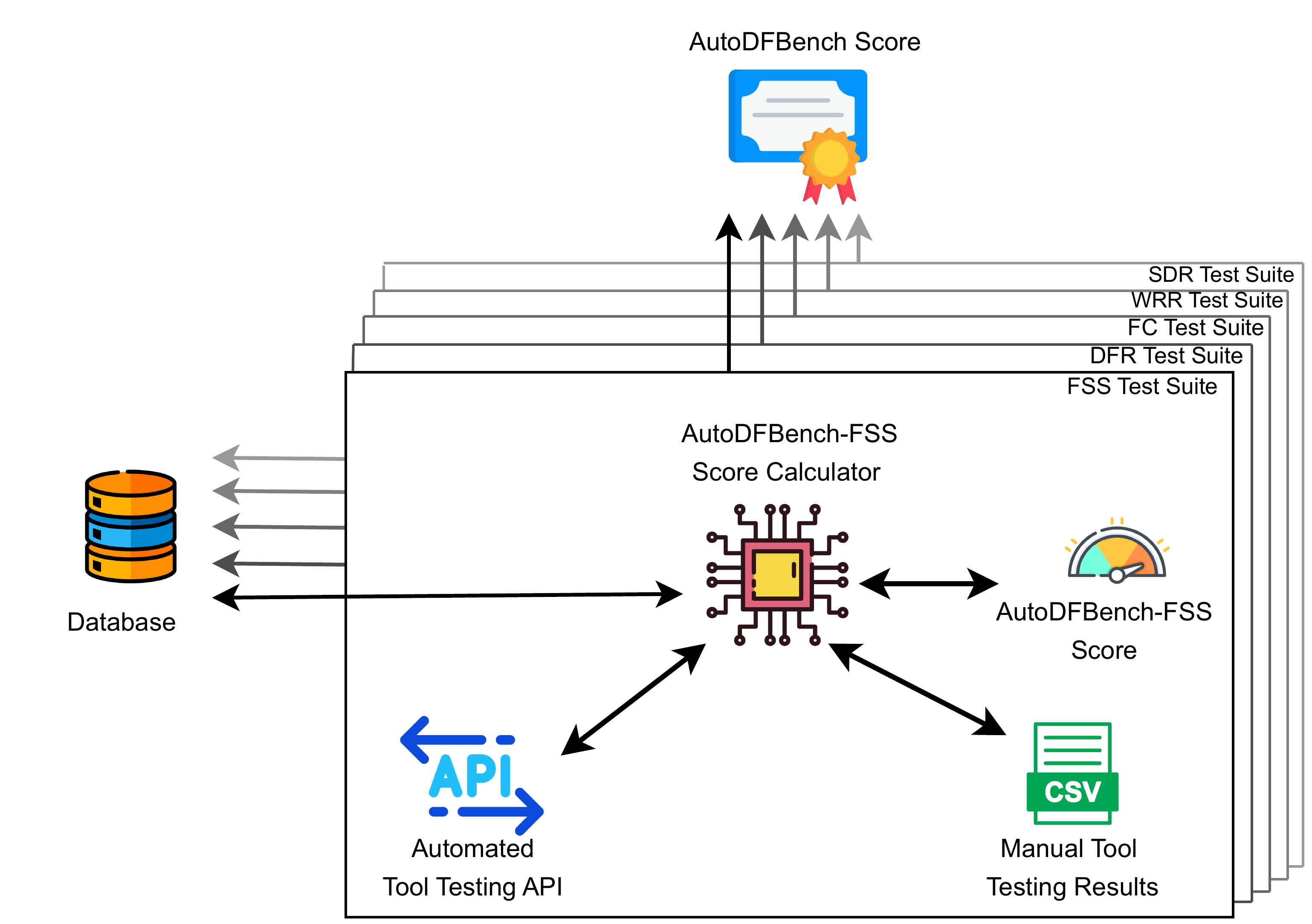}
    \caption{Overview of the proposed framework}
    \label{fig:arch_overview}
\end{figure}

This section outlines the design and architecture of AutoDFBench~1.0 and explains how each component enables reproducible, metrics-driven benchmarking across diverse forensic tasks. The overall design presents the relationship between ground truth data and computed metrics, modular extensibility across test suites, and how ground truth data aligns with NIST CFTT test cases to ensure the comparability and auditability of results.

\subsection{Design Considerations}

The framework adopts a modular architecture in which each evaluation suite operates independently while sharing a common processing layer. Such separation is essential because the evaluation criteria differ across the various forensic tasks covered, i.e., each task (or subtask) may necessitate a particular definition of \textit{correctness} that is unique to that task. While this definition might be reusable across similar tasks, no definition is universal. 

Each test suite functions as an independent module with its own parameters and scoring process, enabling new forensic domains or test categories to be incorporated without requiring any modifications to the existing architecture. The same structure also supports proficiency testing, where independently prepared datasets can be introduced as new modules to assess DF laboratories and tool performance under controlled conditions.

All modules communicate with a central MySQL database that maintains ground truth, configuration, and results to ensure auditability. As illustrated in Figure~\ref{fig:arch_overview}, the system is organised into three layers: \emph{Score Calculation}, \emph{API}, and \emph{CSV Input}. The API layer acts as a mediation interface between tool outputs and scoring logic, enforcing a consistent schema across all suites. The CSV input layer supports batch and offline evaluations, promoting reproducibility and controlled experimentation.

AutoDFBench~1.0 integrates five evaluation suites corresponding to the following DF test suites defined by the NIST CFTT programme: forensic string search, deleted file recovery, file carving, Windows Registry recovery, and SQLite data recovery. Each suite is implemented as a dedicated API endpoint with parameters specific to its test category but serves a shared scoring format that includes precision, recall, and F1~score.

\subsection{Database}

The database schema is structured to maintain minimal redundancy while ensuring evidential completeness. Three primary tables are \texttt{config}, \texttt{ground\_truth}, and \texttt{test\_results}.

The \texttt{config} table contains the configuration parameters required for each module, such as source directories, disk image paths, and environment-specific settings, thereby supporting repeatable experiments across systems.

The \texttt{ground\_truth} table serves as the reference dataset for all evaluations. Each record includes the test case identifier, relevant attributes, and classification fields. The \texttt{type} column differentiates between active, deleted, or unallocated files; \texttt{os} identifies the operating system context; and \texttt{cftt\_task} specifies one of the evaluation domains (\texttt{string\_search}, \texttt{deleted\_file\_recovery}, \texttt{file\_carving}, \texttt{windows\_registry}, or \texttt{sqlite}).  
For deleted file recovery, additional attributes such as \texttt{file\_name}, \texttt{size}, and multiple timestamps (\texttt{access\_time\_stamp}, \texttt{modify\_time\_stamp}, \texttt{change\_time\_stamp}, \texttt{deleted\_time\_stamp}) are included, along with \texttt{dfr\_blocks} for recording original allocation details.  
For file carving tasks, a \texttt{carve\_type} field distinguishes between contiguous (\texttt{contig}), non-contiguous (\texttt{non}), and fragmented (\texttt{frag}) files.

The \texttt{test\_results} table acts as both a results repository and a logging table. Each entry records the test case identifier, the tested tool, counts of true positives, false positives, and the computed F1~score, thereby maintaining a complete audit trail of all evaluations and enabling quantitative comparisons across forensic tools and generated code variants.

\subsection{API Layer}
\label{api_layer}

The API layer provides a standardised interface for evaluation and ensures consistency across test suites while allowing suite-specific parameters. The available endpoints are:
\begin{itemize}
    \item \texttt{/api/v1/string-search/evaluate}
    \item \texttt{/api/v1/deleted-file-recovery/evaluate}
    \item \texttt{/api/v1/file-carving/evaluate}
    \item \texttt{/api/v1/windows-registry/evaluate}
    \item \texttt{/api/v1/sqlite-recovery/evaluate}
\end{itemize}
The string search endpoint requires file content and OS, whereas the deleted file recovery API requires the file name, file size, and MAC timestamps (Modified, Access, Created). For the file carving API, the carved file should be passed as form data. In Windows registry recovery, a converted CSV file is required as the input parameter. For the SQLite recovery API, the database name, file rows, column names, and primary key must be submitted depending on the test case.

Each endpoint also accepts the common parameter base test case, with the tool being used as an optional parameter. All responses return standard metrics, including precision, recall, and the AutoDFBench Score for the specific test case.  

\subsection{CSV Input Layer}
\label{csv_layer}

The CSV input layer extends functionality for batch and offline evaluations, facilitating proficiency studies and automated testing workflows. The execution script \texttt{csv\_eval.py} accepts the test case name, the input CSV file path, and the output report path. The input CSV file should include all relevant test case code and the specific results obtained from the tools.
This mechanism enables bulk processing of test suites without live API calls and ensures that evaluations can be reproduced precisely. The approach supports continuous integration pipelines (CI) and proficiency assessments, where deterministic output generation is critical.

\subsection{Ground Truth Datasets}
\label{Ground_truth}

Ground truth (GT) datasets are directly derived from NIST CFTT resources and include only the attributes necessary to test the mandatory test cases. 

\subsubsection{String Search Ground Truth}
In string search GT, NIST has provided a unique four-digit number for each word match, where this number is concatenated with the string line of the search text. During evaluation, tool-reported string lines are mapped to these identifiers to verify exact matches. For example, in test case \texttt{FT-SS-01}, identifiers \texttt{0896}, \texttt{0898}, and \texttt{0900} indicate the lines that contain the word ``DireWolf'' (the keyword to be searched in these test cases). 

As explained in Section~\ref{intro-fss}, 10 main test cases are defined, which are further divided into sub-test cases representing different keywords, file types, and OS. Overall, there are 1,844 forensic string search sub-test cases.

\subsubsection{Deleted File Recovery Ground Truth}

For DFR test cases, file blocks, file name, file size, access time, modify time, and creation time were used as the ground truth parameters, which were provided by NIST\footnote{\url{https://cfreds-archive.nist.gov/dfr-test-images.html}}. These parameters were maintained for comparison with test cases in the score calculation.

As NIST does not publicly provide the file blocks for files in test case DFR-1, DFR-2 and DFR-3, the file blocks are calculated using the absolute sector numbers concerning the partition offset and the number of sectors that the filesystem allocates per data unit. For each absolute sector \(s\) address, the corresponding file-system block index is calculated as:  

\[
\text{Block}(s) \;=\; \left\lfloor {s - PS} \right\rfloor
\]

where  
\begin{itemize}
  \item \(PS\) is the partition start sector,  
\end{itemize}
With this, the file starting block was identified, and by using the file size and sector size, it was determined how many blocks were needed for that file. With this formula file, blocks are calculated, and block ranges are stored in the ground truth table.

In the DFR test cases DFR-07, DFR-08, DFR-09, DFR-10, and DFR-13, most files lack absolute sectors, which are not included in the GT document. As a result, the GT data are not fully complete for these test cases. Despite this, for all 14 NIST test cases, there are 8,147 ground truth test variations available.

\subsubsection{File Carving Ground Truth}

NIST provided image files in BMP, GIF, PNG, HEIC and TIFF formats, and three disk image types were used as the ground truth data\footnote{\url{https://cfreds-archive.nist.gov/filecarvingtestreports.html}}. Each ground truth test case contains the image file path of each image and the contingency type (three types). In the ground truth table, the base test case is defined with a fragmented type and the image format. For example, the \textit{carve-conti-bmp} test case is related to testing the carved BMP files taken from the disk image of contiguous files. The unique ground truth test variations for file carving, totalling 108, enable the testing of all file carving test cases as defined by NIST.

\subsubsection{Windows Registry Recovery Ground Truth}  

For tool testing, NIST has provided a collection of registry files for each test case. Since they only provided the hive files, it is necessary to extract the content from these files, as the NIST test cases are aligned with the testing of the contents inside the files. To do this, the \texttt{regipy} Python library was used\footnote{\url{https://pypi.org/project/regipy/}}. It allows for reading keys and values inside a hive file. A set of CSV files was generated for each hive file using \texttt{regipy}, and those CSV files were used as the ground truth data to cross-check the keys and values for data sent via the API. \texttt{regipy} was selected because it is a free and open-source library that offers the required capabilities, such as reading subkeys and values. On GitHub, it has gained 261 stars and has active contributors, which verifies the solidity of the library.

This CSV file contains PATH, which maintains the hierarchical registry key namespace, TYPE maintains the data type, VALUE holds the data for a particular key, and MTIME records metadata modification tracking. These CSV files serve as the ground truth data, with the database storing the paths to these extracted files corresponding to each test case. The ground truth data consists of 49 test variations across 28 test cases.

\subsubsection{SQLite Recovery Ground Truth}  

To align with NIST test cases and properly evaluate them, the SQLite recovery ground truth contains a large number of parameters. These parameters include page size, journal mode, number of pages, file hash, encoding type, source file, table names, columns, and number of rows. For each test case, one or more parameters are used as validators. 

For the SFT-01 test case, page size, file hash, journal mode, and number of pages act as the verification parameters. For SFT-02, table names, column names, and row ID counts are taken as the ground truth data. SFT-03 maintains the database file name and row count, and for SFT-04, it retains the source file path as the ground truth data.

Ground truth data were extracted from the NIST CFTT SDR files. To extract the necessary data from the SQLite file, \textit{DB Browser for SQLite} was used\footnote{\url{https://github.com/sqlitebrowser/sqlitebrowser}}. This was chosen because it is a popular open-source tool on GitHub. This ground truth dataset includes 4 test cases and 820 variations of test data. 

\subsection{Score Calculation}
\label{ScoreCalculation}

Score calculation constitutes the analytical core of the framework, transforming raw evaluation outputs into reproducible and comparable performance metrics. Each evaluation suite applies distinct scoring logic tailored to the forensic characteristics of its test category while maintaining a common set of quantitative indicators: precision, recall, and F1~score. These metrics are first computed at the subtest level, corresponding to individual CFTT cases (e.g., FT-SS-01 or DFR-01), and are then averaged to obtain a consolidated test suite score, defined as the AutoDFBench test suite score. The average of all five test suite scores forms the AutoDFBench Score, which represents the overall accuracy of a tool.

By employing this hierarchical scoring structure, the framework ensures that every evaluation, from granular subtests to complete benchmark suites, can be interpreted, replicated, and compared in a statistically consistent manner.

\subsubsection{Forensic String Search Score}

In the string search score, evaluation is based on mapping the returned string line identifiers to the ground truth. Correct matches are counted as true positives (TP), additional lines as false positives (FP), and missed identifiers as false negatives (FN). These values are used to compute precision, recall, and the F1~score for each test case, and then the Average for all the test scores as the AutoDFBench-FSS score.

\subsubsection{Deleted File Recovery Score}

Score calculation in DFR varies between test cases for overwritten (DFR-07, DFR-08, DFR-10) and non-overwritten test cases (DFR-01, DFR-02, DFR-03, DFR-05, DFR-06, DFR-09, DFR-12). For these, only deleted file blocks are considered, and it validates how many complete file blocks match the GT record. As a result, a TP is given if a recovered file matches its full set of constituent blocks. If the recovered file blocks do not contain all the blocks, it counts as a FP. If the file blocks are not relevant to a particular ground truth entry, it is counted as a FN. Subsequently, the precision, recall, and F1~score are calculated.

For the MAC times recovery test cases (DFR-01-MAC), the correct MAC times are validated, and only when all timestamps match is this considered a TP. Similarly, the file size evaluation test cases (DFR-07-SIZE, DFR-01-SIZE, DFR-011-SIZE) verify that the correct file size is captured and counted as a TP. The Latin character file test (DFR-04-CHAR) validation counts a TP when an exact file name is matched. FP is determined when a parameter does not match the ground truth, and FN is determined when the ground truth record is not coherent with the parameter record.

The F1~scores are then calculated and averaged across all the test cases to produce the AutoDFBench-DFR score.

\subsubsection{File Carving Score}  
The file carving score is calculated using the same principle as deleted file recovery by determining TP, FP, and FN to compute precision, recall, and the F1~score. Each carved file is compared against the ground truth by analysing its byte sequence while also verifying that the file is decodable and can be opened, as this is essential for its use as a forensic artefact.  

To identify the most relevant carved output matched to each ground truth file, the framework employs a perceptual hash (\emph{pHash}) to select the closest visual match. Perceptual hashing generates a compact representation of an image based on its visual features, such that similar images produce hashes with a small Hamming distance, even if they differ at the pixel level due to compression or minor alterations. This robustness makes pHash well suited for validating recovered images in forensic scenarios~\citep{SAMANTA2021203}.

 The similarity between the carved file \(f_c\) and the ground truth file \(f_g\) is measured at the byte level. If the similarity is greater than or equal to 20\% and the carved file is decodable, the recovery is counted as a TP. Otherwise, if the tool provides a carved file that does not correspond to any ground truth file, it is counted as a false positive (FP). Conversely, if a ground truth file has no corresponding carved output, or does not meet the similarity threshold, it is counted as a FN. With TP, FP and FN, the F1~score is calculated, and the Average F1~score is defined as the AutoDFBench-FC score

The 20\% threshold was determined by carving the NIST-provided test image set, using Scalpel v1.6, where files that were empirically assessed as usable each exceeded 20\% byte similarity.

\subsubsection{Windows Registry Recovery Score}
Similar to string search calculations, the Windows registry uses the same mechanism for text comparison. As the ground truth CSV file contains the path and value for each test case, these values are cross-checked at the string level.

When the row matches the hive file CSV, which is recovered by the tool, it is counted as TP. When the ground truth CSV contains a row that is not included in the CSV sent by the recovered tool, it is counted as a FN. Additionally, FP is counted when the recovered tool CSV transmits a row that is not present in the ground truth CSV file.
With this data framework, the precision, recall, and F1~score are calculated.

\subsubsection{SQLite Recovery Score}

SQLite score calculation uses a mechanism similar to that of all other score calculators. However, for each test case, the attributes used for score calculation differ. SFT-01 checks the page size, journal mode, number of pages, file hash, and text encoding. The correct matching of these parameters counts toward the true positives (TPs) for SFT-01. If the passed parameter does not match the ground truth, it is counted as a FP, while a FN results in a zero value. This is due to the fact that SQLite recovery test cases validate a true or false condition with the given parameters from the API.

In the SFT-02 table, names, column names, and row counts are compared. In this test case, all column names, table names, and row counts must match exactly to be counted as a TP. Similar to the SFT-01, a FP is enumerated for scenarios where the ground truth data does not match. Moreover, the FN count remains zero.
SFT-03 uses deleted and updated row IDs to check the ground truth data. Correctly matching row IDs are counted as a TP, while a FP counts rows that exist in the ground truth but are not included in the API request. A FN is counted for rows for which the ground truth does not have a valid record pertaining to that comparison.
SFT-04 keeps the SQLite file name as the cross-checker, and if the correct file name matches, it is counted as a TP, and if not, as a FP, and again, a FN is counted as zero.

F1~scores are computed separately for each test case, and the average of each test variation is then defined as the AutoDFBench-SDR score.

\section{Experimentation}
\label{ExperementFlow}

To evaluate the integrity and accuracy of the framework APIs and the score calculation functionality, experiments were conducted using the publicly available test datasets provided by NIST. Since each test suite contains a distinct set of test cases, the experiments were executed individually for each suite. The framework was tested in a Conda environment, where all necessary Python libraries were installed to support the APIs for each test suite. The corresponding libraries, along with API requests and responses, are made available in the GitHub repository referenced in Section~\ref{intro}.

\subsection{Forensic String Search Experimentation}  

The FSS test data were obtained from NIST's Federated Testing v4.0 files. These datasets include the exact string lines expected as output for each test case, together with the corresponding disk image (\texttt{*.dd}) files for both Linux and Windows environments\footnote{\url{https://cfreds.nist.gov/all/NIST/StringSearch,V11}}.
For evaluation, the provided string lines were submitted to the framework via the API, along with the corresponding base test case name. To automate this process, all string line parameters were defined in a CSV file and mapped to each test case. A Python script was then used to execute the test cases sequentially, storing the output F1~scores in another CSV file.

\subsection{Deleted File Recovery Experimentation}

To verify the file block matching approach for the ground truth data DFR-01, DFR-02, DFR-03, and DFR-05  test cases, tests were conducted using The Sleuth Kit 3.2.2 (TSK~3.2.2)\footnote{\url{https://www.sleuthkit.org/sleuthkit/history.php}}. For this, NIST has provided disk image files for each file system, i.e., EXT, FAT, and NTFS. Using these image files, deleted records were observed, and the file blocks for each test case were recorded. They were then tested via the framework's API. Due to the unavailability of data for ground truth, DFR-06, DFR-07, DFR-08, DFR-09, DFR-10, and DFR-013 test cases were not evaluated.

Recovery MAC times and deletion through the recycle bin were also tested as a subtest case of DFR-01. Similarly, file size evaluation scenarios were tested as subtest cases of DFR-01 and DFR-11. NIST uses these subtest cases because they utilise the same disk image files for those tests. However, the main purpose of the experimentation is to validate the logic and score calculation of the framework, enabling the testing of these test cases using data from the NIST documentation\footnote{\url{https://cfreds.nist.gov/all/NIST/DeletedFilesRecovery}}.

\subsection{File Carving Experimentation}

As the scope of the experimentation in this paper is to determine the correctness of the scoring mechanism and the validity of ground truth data, the file carving scenario was tested with the same images provided by NIST that were used in the ground truth data. According to the NIST carving test cases, contiguous file scenarios are FC-01, FC-02 and FC-03, fragmented file scenarios are FC-04 and FC-05, and FC-05 represents the non-aligned cluster scenario. FC-07 was not tested, as no related NIST data is available for ground truth to check the partial file.

\subsection{Windows Registry Recovery Experimentation}

Since the Windows registry ground truth was harvested by a Python module, validating the ground truth data and the score calculation mechanism is necessary. To perform this validation, another Python registry extraction module called \texttt{python-registry} was used. 
Using these Python libraries, each of the NIST-provided hive files was extracted and saved as separate CSV files.
To validate the ground truth data and the API, the CSV files created by \texttt{Regipy} were also sent to the API. Output F1~scores were recorded from the CSV files generated by both tools.

For each test case, the tool-generated CSV was submitted to the Windows Registry Recovery API endpoint with the corresponding parameters, including the base test case identifier, tool designation, and job identifier. The API executed entry-level comparisons through normalised key matching, generating performance metrics including true positives, false positives, false negatives, precision, recall, and F1~scores.

\subsection{SQLite Recovery Experimentation}

The dataset was first downloaded from NIST, which contains all SQLite files covering the core test cases.
These files were then loaded into the DB Browser for SQLite, and the required parameters were extracted to send to the API. In SFT-01, database headers such as \emph{journal mode} and \emph{page size} were obtained using the PRAGMA \emph{page\_size; PRAGMA journal\_mode;}. Similarly, in SFT-02, table names, column names, and row counts were extracted. In SFT-03, the modified and deleted rows were identified using the primary keys, as explicitly documented in the SQLite test case descriptions provided by NIST. These identifiers were directly passed to the API. SFT-04 was tested by sending the exact SQL file name via the API.

\section{Results and Discussion}
\label{ResultsDiscussion}

This section presents the results of the experiments conducted to validate the framework and examines the AutoDFBench Scores obtained for each test suite. The discussion highlights the accuracy, consistency, and reproducibility of the framework’s evaluations.

\subsection{Forensic String Search Results}
\label{FSSResults}

Via the result CSV files, all sub-test case F1~scores are averaged into the main test cases, as represented in Table~\ref{tab:fssresults}. For all test cases, the F1~score of 1 illustrates the validity of the ground truth and the correctness of the framework's score calculation.

\begin{table}[htbp]
\centering
\small
\caption{Forensic String Search Test Results}
\label{tab:fssresults}

\begin{tabular}{|c|c|}
\hline
\textbf{Test Case} & \textbf{F1~score} \\
\hline
FT-SS-01 & 1\\
\hline
FT-SS-02 & 1 \\
\hline
FT-SS-03 & 1 \\
\hline
FT-SS-04 & 1 \\
\hline
FT-SS-05 & 1 \\
\hline
FT-SS-06 & 1 \\
\hline
FT-SS-07 & 1 \\
\hline
FT-SS-08 & 1 \\
\hline
FT-SS-09 & 1 \\
\hline
FT-SS-10 & 1 \\
\hline
\end{tabular}

\end{table}

\subsection{Deleted File Recovery Results}
\label{DFRResults}

Table~\ref{tab:dfr_testcasesresults} presents the results of the DFR experiments. Test cases marked with `–' denote that they were ignored for testing due to gaps in data availability. Each of the evaluated test cases obtained an F1~score of 1, which demonstrates the validity of the score calculation logic, block calculation logic, and the ground truth data. 

\begin{table}[h!]
\centering
\small
\caption{Deleted File Recovery Test Results}
\label{tab:dfr_testcasesresults}
\resizebox{\columnwidth}{!}{%

\begin{tabular}{|c|c|c|}
\hline
\textbf{Test Case} & \textbf{\makecell{F1~score with \\ NIST Test Data}} & \textbf{\makecell{F1~score \\ with TSK 3.2.2}} \\ \hline
DFR-01 &   & 1 \\ \hline
DFR-01  Recovered MAC Times &  1 &  \\ \hline
DFR-01  Deletion Through Recycle  &  1 &  \\ \hline
DFR-01  File Size   & 1  &  \\ \hline
DFR-02 &   & 1 \\ \hline
DFR-03 &   & 1 \\ \hline
DFR-04 & 1  &  \\ \hline
DFR-05 & 1 &  \\ \hline
DFR-06 & - &  -\\ \hline
DFR-07 & -  & - \\ \hline
DFR-07 File Size & 1  & - \\ \hline
DFR-08 & -  & - \\ \hline
DFR-09 & -  & - \\ \hline
DFR-10 & -  & - \\ \hline
DFR-11 & 1  &  \\ \hline
DFR-11 Special NTFS Situations & 1  &  \\ \hline
DFR-11 File size & 1  &  \\ \hline
DFR-12 & 1  &  \\ \hline
DFR-13 & -  & - \\ \hline
DFR-14 & 1  &  \\ \hline

\end{tabular}
}
\end{table}

\subsection{File Carving Results}
\label{fcResults}

As shown in Table~\ref{tab:fcresults}, for all contiguous file scenarios (FC-01, FC-02, FC-03), the F1~score was 1, as was the case for similarly fragmented scenarios (FC-04, FC-06) and the non-aligned cluster scenario (FC-06). `–' denotes that FC-07 was not tested. Results show that all the test cases obtained an F1~score of 1, which indicates the correctness of the score calculation and the file matching logic.

\begin{table}[h!]
\centering
\small
\caption{File Carving Test Results}
\label{tab:fcresults}

\begin{tabular}{|c|c|}
\hline
\textbf{Test Case} & \textbf{F1~score} \\ \hline
FC-01 & 1 \\ \hline
FC-02 & 1 \\ \hline
FC-03 & 1 \\ \hline
FC-04 & 1 \\ \hline
FC-05 & 1 \\ \hline
FC-06 & 1 \\ \hline
FC-07 & - \\ \hline
\end{tabular}
\end{table}

\subsection{Windows Registry Recovery Results}
\label{WRRResults}

Table~\ref{tab:wrr_testcases_results} summarises the WRR experimental results. As NIST provides only the binary hive images and not the corresponding extracted registry content, the \texttt{regipy} and \texttt{python-registry} libraries were used to generate and evaluate ground truth datasets. 

For regipy, the F1~score is 1 for all the test cases, as it was used to create the ground truth data. However, for python-registry, the F1~score is near 1 in most test cases. There are still some test cases that have very low values for the data from the Python registry library. 
Both of these comparison values demonstrate the validity of the ground truth data and the score calculation logic.

F1~scores were computed for each test case by averaging the results of all its sub-cases. Test cases marked with `–' indicate that extraction was not possible due to dataset gaps or library processing failures, preventing the calculation of a corresponding F1~score.

\begin{table}[h!]
\centering
\small
\caption{Windows Registry Recovery Test Results}
\label{tab:wrr_testcases_results}
\begin{tabular}{|c|c|c|}
\hline
\textbf{Test Case} & \textbf{\makecell{F1~score\\for regipy}} & \textbf{\makecell{F1~score for\\python-registry}} \\ \hline
CR-01 & - & - \\ \hline
CR-02 & 1 & 0.9509 \\ \hline
CR-03 & 1 & 0.9509 \\ \hline
CR-04 & 1 & 0.6074 \\ \hline
CR-05 & - & - \\ \hline
MR-01 & 1 & 0.0036 \\ \hline
MR-02 & 1 & 0.9509 \\ \hline
MR-03 & 1 & 0.9556 \\ \hline
MR-04 & 1 & 0.9553 \\ \hline
MR-05 & 1 & 0.9509 \\ \hline
MR-06 & 1 & 0.9509 \\ \hline
MR-07 & 1 & 0.9509 \\ \hline
MR-08 & 1 & 0.9509 \\ \hline
MR-09 & 1 & 0.9509 \\ \hline
MR-10 & 1 & 0.9509 \\ \hline
MR-11 & 1 & 0.9509 \\ \hline
MR-12 & 1 & 0.9509 \\ \hline
MR-13 & 1 & 0.9509 \\ \hline
MR-14 & 1 & 0.9509 \\ \hline
MR-15 & 1 & 0.9283 \\ \hline
NR-01 & 1 & 0.25 \\ \hline
NR-02 & 1 & 0.7273 \\ \hline
NR-03 & 1 & 0.999 \\ \hline
NR-04 & 1 & 0.7901 \\ \hline
NR-05 & 1 & 0.7273 \\ \hline
NR-06 & 1 & 0.4222 \\ \hline
NR-07 & 1 & 0.4706 \\ \hline
NR-08 & - & - \\ \hline

\end{tabular}
\end{table}

\subsection{SQLite Data Recovery Results}
\label{SDRResults}

Table~\ref{tab:sdr_results} presents the results of the SQLite data recovery experiments. For all core test cases, the F1~score is 1. The purpose of the experiment is to validate the accuracy of the ground truth data and the score calculation function by testing the results. With the SDR test suites, it is demonstrated that the ground truth is complete and that the framework functions as expected.

\begin{table}[h!]
\centering
\small
\caption{SQLite Data Recovery Results}
\label{tab:sdr_results}
\resizebox{0.4\columnwidth}{!}{%

\begin{tabular}{|c|c|}
\hline
\textbf{Test Case} & \textbf{F1~score}  \\ \hline
SFT-01 & 1  \\ \hline
SFT-02 & 1 \\ \hline
SFT-03 & 1   \\ \hline
SFT-04 & 1   \\ \hline

\end{tabular}
}
\end{table}

\section{Discussion}
\label{Discussion}

The results obtained from all test suites demonstrate that the framework performs as expected, producing consistent and accurate evaluation outcomes. In the Deleted file recovery and Windows registry test cases, a few ground truth datasets were incomplete or unavailable, which limited the ability to fully validate certain scenarios, as discussed in Section~\ref{DFRResults} and Section~\ref{WRRResults}. Despite these limitations, all other test cases across the five evaluated test suites produced the expected outputs, confirming the validity and integrity of the framework.

In contrast, test suites with deterministic outputs, such as string search, file carving, Windows registry recovery, and SQL data recovery, achieved perfect F1~scores as the match sets were precise and the ground truth data were well defined.

The benchmarks' trustworthiness is measured by the reproducibility of each test. Of course, in any digital forensics context, reproducibility is a necessity. This framework promotes reproducibility, transparency, and auditability, making it a reliable benchmark for the digital forensics community. 
It also adheres to a deterministic scoring logic, which is another key quality of an accurate benchmarking framework.
Open-source access to the code and ground truth enables verification and extension of the framework in the future. The modular architecture is another key advantage, as it helps maintain the base architecture and facilitates the addition of more modules. On top of ground truth data and score calculation logic, this framework can be extended for practitioner proficiency testing and to evaluate the performance of future DF AI agents. 

Several limitations were identified during the experimentation. The absence of complete ground truth data from NIST datasets prevents full verification of certain scenarios. This might be perceived as a drawback of the framework. However, the problem arises not because of the framework but due to the lack of accessibility to ground truth data.
Additionally, the current version of the framework focuses mainly on quantitative output comparison, such as byte- or block-level matching or string-level matching. It does not yet evaluate the meaning or context of the results. For example, it cannot assess the percentage of a carved file that was carved or the number of blocks recovered from a file. As a result, it cannot distinguish between minor acceptable variations and genuine forensic errors. 
In the current scope of the framework, it considers only the core evaluation test cases. However, it is also vital to include optional test cases and their ground truth in future work. Additionally, the framework does not measure the processing speed or resource usage of the tested tools, meaning that performance and efficiency aspects were not considered in this benchmark. However, it can be reasonably assumed that accuracy is the most important characteristic in any digital forensic context.

Overall, these findings demonstrate that AutoDFBench~1.0 offers a reliable and reproducible approach for validating digital forensic tools and AI-generated code, while highlighting areas where further refinement and human oversight will strengthen the framework’s forensic robustness.

\section{Future Work}
\label{future}

The modular architecture of AutoDFBench enables seamless extension to additional NIST test suites and emerging forensic domains, including memory forensics, mobile device forensics, and cloud data extraction. Future work will focus on integrating these new test categories and expanding the database of ground truth datasets to enhance coverage and accuracy.  

Furthermore, the framework demonstrates strong potential for use in proficiency testing and certification exercises, enabling forensic laboratories and practitioners to systematically assess tool performance and examiner competency under standardised conditions. In the long term, AutoDFBench could serve as the foundation for a universal benchmarking and validation platform that supports both academic research and operational digital forensic practices.

\section{Conclusion}
\label{Conclusion}

This paper presents AutoDFBench~1.0, an extended and enhanced version of AutoDFBench that serves as an automated, modular, and extensible benchmarking framework for evaluating both conventional and AI-generated digital forensic tools. Built upon the principles of the NIST CFTT programme and NIST ground truth data, the framework standardises the evaluation process across multiple forensic scenarios, including string search, file carving, deleted file recovery, Windows Registry analysis, and SQLite data recovery.  

Experimental validation using NIST datasets demonstrated that the framework produces accurate, consistent, and reproducible results. By quantifying performance through precision, recall, and F1~score, AutoDFBench establishes a transparent and repeatable approach to validating forensic tools and generated code. The framework thereby contributes to improving trust, standardisation, and scientific integrity within the digital forensics community.

\bibliographystyle{elsarticle-harv}

\bibliography{sample-base}

\end{document}